# TESTING INFLATION WITH THE COSMIC BACKGROUND RADIATION

J. Richard BOND
*CIAR Cosmology Program,*
*Canadian Institute for Theoretical Astrophysics,*
*University of Toronto, Toronto, Ontario, CANADA*

## Abstract

In inflation cosmologies, cosmic structure develops through the gravitational instability of the inevitable quantum noise in primordial scalar fields. I show how the acceleration of the universe defines the shape of the primordial spectrum of gravitational metric and scalar field fluctuations. I assess how we can determine the shape and overall amplitude over the five decades or so of spatial wavelengths we can probe, and use current data from cosmic background radiation (CMB) anisotropies, large scale clustering and streaming observations, distribution functions of cosmic objects, to show how far we are in this program. Broad-band power amplitudes are given for CMB anisotropy detections up to spring 1994, covering angular scales from all-sky down to arcminutes: DMR, FIRS, Tenerife, SP91, BigPlate, Python, SP89, MAX, MSAM, ARGO, White Dish, OVRO. It may be only a little premature to say that a spectral shape is emerging which is near that of preferred inflation models. The cluster-scale density fluctuation power derived from the broad-band power for COBE must fall within a narrow range to get the abundance of clusters right. This rules out many structure formation models, in particular restricting the primordial spectral index to be close to the scale invariant one. I show that COBE band-powers found with full Bayesian analysis of the $53, 90, 31$ $a+b$ GHz first year DMR (and FIRS) maps are in good agreement, and are essentially independent of spectral slope and degree of (sharp) signal-to-noise filtering. Further, after (smooth) optimal signal-to-noise filtering (*i.e.,* Weiner-filtering), the different DMR maps reveal the same large scale features and correlation functions with little dependence upon slope. However, the most probable slope depends upon how the maps are filtered: with no filtering whatsoever, the slope is high, but the power is not described by a single-slope law; as filtering is increased, the index moves nearer to inflation predictions.[1]

---

## 1. Introduction

The inflation paradigm — that the region of the Universe in which we reside was once in a state of accelerated expansion — remains the best way to account for local homogeneity and isotropy. And, of course, quantum noise generated during acceleration is a natural byproduct that may account for the observed structure within our Hubble patch. The data on the cosmic background and large scale cosmic structure is rapidly approaching a state where this can be tested in detail. I discuss the overall issue, describing and extending post-DMR detection work in [1]-[7].

### 1.1. Fluctuation Variables and Their Power Spectra

Over the length scales we can probe with observations, *i.e.*, within our "Hubble patch", the post-inflation distribution of the noise seems most likely to be linear fluctuations on a slowly varying background geometry.

A generic fluctuation variable $D(\mathbf{x},t)$ can be expanded in terms of modes $\mathcal{M} \in \{$ adiabatic scalar, isocurvature scalar, vector or tensor; growing or decaying $\}$:

$$D(\mathbf{x},t) = f \sum_{\mathbf{k}\mathcal{M}} \left\{ u^{(D)}_{\mathbf{k}\mathcal{M}}(t) Q_{\mathbf{k}\mathcal{M}}(\mathbf{x}) a_{\mathbf{k}\mathcal{M}} + u^{(D)*}_{\mathbf{k}\mathcal{M}}(t) Q^*_{\mathbf{k}\mathcal{M}}(\mathbf{x}) a^\dagger_{\mathbf{k}\mathcal{M}} \right\} \qquad (1)$$

$$f = 1/2 \text{ classical}, \quad f = 1 \text{ quantum}.$$

For classical fluctuations, $a_{\mathbf{k}\mathcal{M}}$ is a random variable and $a^\dagger_{\mathbf{k}\mathcal{M}}$ its complex conjugate, while for quantum fluctuations, $a_{\mathbf{k}\mathcal{M}}$ is an annihilation operator for the mode $\mathbf{k}\mathcal{M}$ and $a^\dagger_{\mathbf{k}\mathcal{M}}$ is the creation operator. The $u^{(i)}_{\mathbf{k}\mathcal{M}}(t)$ are mode functions which describe the evolution. The spatial dependence of the modes is given by eigenfunctions $Q_{\mathbf{k}\mathcal{M}}(\mathbf{x})$ of the Laplacian of the background geometry. For a flat background of most relevance to inflation models, it is simply a plane wave, $Q_{\mathbf{k}\mathcal{M}}(\mathbf{x}) = e^{i\mathbf{k}\cdot\mathbf{x}}$, labelled by a comoving wavevector $\mathbf{k}$. For curved backgrounds, the eigenfunctions are more complex.

The power spectrum of $D$ associated with mode $\mathcal{M}$ is the fluctuation variance per log wavenumber and can be expressed in terms of the statistics of $a_{\mathbf{k}\mathcal{M}}$ and $a^\dagger_{\mathbf{k}\mathcal{M}}$:

$$\text{quantum}: \quad \mathcal{P}_{D|\mathcal{M}}(k) \equiv \frac{d\sigma^2_{D|\mathcal{M}}}{d\ln k} = \frac{k^3}{2\pi^2} |u^{(D)}_{\mathbf{k}\mathcal{M}}(t)|^2 \left(1 + 2\langle a^\dagger_{\mathbf{k}\mathcal{M}} a_{\mathbf{k}\mathcal{M}} \rangle \right), \qquad (2)$$

$$\text{classical}: \quad \mathcal{P}_{D|\mathcal{M}}(k) = \frac{k^3}{2\pi^2} |u^{(D)}_{\mathbf{k}\mathcal{M}}(t)|^2 \langle a^*_{\mathbf{k}\mathcal{M}} a_{\mathbf{k}\mathcal{M}} \rangle. \qquad (3)$$

If the modes are Gaussian-distributed, statistically homogeneous and isotropic, then this is all that is needed to specify the patterns in the field $D(\mathbf{x},t)$.

In the inflation picture, the wavenumbers in the observable regime are usually considered to be so high that any pre-inflation mode occupation, $\langle a^\dagger_{\mathbf{k}\mathcal{M}} a_{\mathbf{k}\mathcal{M}} \rangle$, is negligible, and only the unity zero point oscillation term appears. In that case, we connect to the random field description by making the real and imaginary parts of $a_{\mathbf{k}\mathcal{M}}$ Gaussian-distributed with variance $1/2$. Although quantization is at least self consistent in linear perturbation theory about a classical background, there are still obvious subtleties associated with the transition from a quantum to a classical random field description. A true inconsistency appears if we include the nonlinear backreaction of the fluctuations upon the background fields and upon themselves. For this, we would need a quantum gravity theory. The stochastic inflation theory is an attempt to bypass this, by treating the fluctuations quantum-mechanically and the inhomogeneous background classically, with the fluctuations influencing the background through stochastic noise terms in a network of Langevin equations for the field variables, *e.g.*, [8, 9].

Over the observable $k$-range, it is convenient to separate the issues of **overall amplitude** for $\mathcal{P}_{D|\mathcal{M}}(k)$ — characterized say by $\mathcal{P}_{D|\mathcal{M}}(R_n^{-1})$ at some normalization length scale $R_n$, or, better, by an integral *wrt* a filter, $\int W(kR_n)\mathcal{P}_{D|\mathcal{M}}(k)d\ln k$, from **shape** — characterized by an index

$$n_D(k) + 3 \equiv d\ln \mathcal{P}_{D|\mathcal{M}}(k)/d\ln k . \tag{4}$$

Thus $-n_D$ is a "fractal dimension": zero is white noise, while three is scale invariance in $D$, or flicker noise, with each octave contributing the same loudness.

In the inflation regime,

$$D \in \{\delta\phi_{inf}, \delta\phi_{is}, h_+, h_\times, \delta\ln a, \delta\ln H, \delta q, \ldots\} . \tag{5}$$

That is, $D$ would refer to fluctuations in (1) the inflaton field $\delta\phi_{inf}$ whose equation of state can give the negative pressure needed to drive the acceleration, (2) other scalar field degrees of freedom $\delta\phi_{is}$ which can, for example, induce scalar isocurvature perturbations,[2] (3) gravitational wave modes $h_+, h_\times$, (4) the inhomogeneous scale factor $a(\mathbf{x},t)$, Hubble parameter $H(\mathbf{x},t)$ and

$$\text{deceleration parameter}: \quad q(\mathbf{x},t) \equiv -d\ln Ha/d\ln a , \tag{6}$$

encoding scalar metric perturbations and their variations. Inflation ends when $q$ passes from negative to positive. Provided the fluctuations over the observable $k$-range remain Gaussian, the outcome of inflation is therefore a set of amplitudes for scalar metric (adiabatic) perturbations, gravity wave modes and various possible isocurvature modes, and primordial spectral index functions for each, in particular:

$$\text{scalar}: n_s(k) \equiv 1 + \frac{d\ln \mathcal{P}_{\ln a|_H}(k)}{d\ln k} , \text{ where } \delta\ln a|_H \equiv \delta\ln a(\mathbf{x}, t(\mathbf{x}, H^{-1})) , \tag{7}$$

$$\text{tensor}: n_t(k) \equiv -3 + \frac{d\ln \mathcal{P}_{GW}(k)}{d\ln k} , \text{ where } \mathcal{P}_{GW}(k) \equiv \mathcal{P}_{h_+}(k) + \mathcal{P}_{h_\times}(k) . \tag{8}$$

Measuring the power in scalar metric fluctuations on the time surfaces upon which the inhomogeneous Hubble parameter $H(\mathbf{x},t)$ — the proper time derivative of $\ln a(\mathbf{x},t)$ — is constant is useful [11, 12, 13, 9]: Once $Ha$ exceeds $k$ for a mode with wavenumber $k$, it becomes time-independent during an inflation epoch with a single dynamically-important scalar field, and it remains so through reheating and the passage from radiation into matter dominance until $Ha$ falls below $k$ (the wave "re-enters" the horizon). [3]

In the post-inflation period,

$$D \in \{\delta\rho_{cdm}, \delta v_{cdm}, \delta\rho_B, \delta v_B, \delta f_\gamma, \delta f_{er\nu}, \delta f_{m\nu}, h_+, h_\times, \Phi, \ldots\} . \tag{9}$$

That is, $D$ would refer to fluctuations in the density and velocity of dark matter and baryons ($\delta\rho_{cdm}, \delta v_{cdm}, \delta\rho_B, \delta v_B$), in the distribution functions for photons ($\delta f_\gamma$) and relativistic or semi-relativistic neutrinos ($\delta f_{er\nu}, \delta f_{m\nu}$), and in the metric (dispersing gravitational wave modes $h_{+,\times}$ and the gravitational potential for scalar fluctuations $\Phi = -\delta \ln a$). The Gaussian nature of the statistics is not modified until mode-mode coupling occurs in the nonlinear regime.

---

2   If axions are the dark matter, $\phi_{is}$ would be the axion field. The isocurvature baryon mode would need to have a $\phi_{is}$ coupled some way to the baryon number, *e.g.*, [16].

3   To be precise about the scalar perturbation quantities used in practice, in the notation of Bardeen [10], $\delta\ln a = \Phi_H$, $\delta\ln H = \bar{H}^{-1}\dot{\Phi}_H + \Phi_H$ in the 'longitudinal gauge', where $H$ is defined as $-1/3\times$ 'trace of the extrinsic curvature'. A translation to the time surface on which $\delta H = 0$ gives $\delta\ln a|_H$, which [9] used to characterize the metric amplitudes in stochastic inflation. But in linear perturbation theory, $\delta\ln a|_H = \delta\ln a - \frac{d\ln a}{d\ln H}\delta\ln H$ is just Bardeen's $\varphi_{com}$ (where $-\nabla^2\varphi_{com}/(4\bar{a}^2)$ is the 3-curvature on comoving hypersurfaces). It is related to $\zeta_{bst}$ of [11] by $\zeta_{bst} = \varphi_{com} + \nabla^2\Phi_H/(3(Ha)^2(1+q))$. The latter term is small when $k < Ha$: both are nearly constant 'outside the horizon' as long as the inflation models are not too outrageous (see § 2.). The use of $\varphi_{com}$ was advocated by [13, 15] and $\zeta_{bst}$ by [11, 12, 14]. Either will do.

The goal of much of cosmology is to use observations of structure in our Hubble patch to piece together the power spectra for observables, then from these to infer the power spectra for the post-inflation fluctuations, $i.e.$,

$$\{\mathcal{P}_{\ln a|_H}(R_n^{-1}), \mathcal{P}_{GW}(R_n^{-1}), \mathcal{P}_{\delta\phi_{is}}(R_n^{-1}), n_s(k), n_t(k), n_{is}(k)\} , \qquad (10)$$

and thereby learn about the physics of the early universe. Hampering this program is the large number of unknown cosmological parameters. We know well the CMB temperature $T_{cmb} = 2.726 \pm 0.005$ [17] and the number of light relic neutrinos, hence $\Omega_\gamma$ and, apparently, $\Omega_{er\nu}$. We do not know the 'global' parameters

$$\{h, \Omega_B, \Omega_\Lambda, \Omega_{cdm}, \Omega_{hdm} \equiv \Omega_{m\nu}, \Omega_{curv} \equiv 1 - \Omega_{tot}, \ldots\} , \qquad (11)$$

as well as energy densities and lifetimes for any decaying particles that were once present. (Here h is the Hubble parameter in units of 100 km s$^{-1}$ Mpc$^{-1}$ and the density parameters are $\Omega_j \equiv \rho_j/\rho_{cr}$, where $\rho_{cr} = 10.5$ h$^2$ kev cm$^{-3}$.) A reasonably strong case can be made that we actually know $\Omega_B h^2 = 0.0125$ to within 10% or so [18]. The small curvature fluctuations observed with COBE is suggestive of small mean curvature, $\Omega_{curv} \ll 1$. $\Omega_\Lambda \approx 0$ is preferred over the odd physics that would be required to make $\Omega_\Lambda$ (or $\langle V(\phi)\rangle/(3 \times 10^{-12}\text{GeV})^4$) significant just at the current time. The favoured theoretical hypothesis is then that the total density in $non$-$relativistic$ matter, $\Omega_{nr}$, is 1, but, with the best astronomical values for h ($\sim 0.7 - 0.8$), one gets a globular cluster age crisis unless $\Omega_\Lambda$ or $\Omega_{curv}$ is nonzero — or something else exists whose energy density varies more slowly than the $\bar{a}^{-3}$ of nonrelativistic matter.

As well as the unknowns in 'global' parameters, astrophysical functions required for mapping from observable to inflation spectra are unknown. Examples are biasing factors relating power spectra for galaxies ($\mathcal{P}_{gg}$), clusters ($\mathcal{P}_{cc}$), $etc.$ to those for the underlying mass density field ($\mathcal{P}_{\rho\rho}$),

$$b_g^2(k) \equiv \mathcal{P}_{gg}(k)/\mathcal{P}_{\rho\rho}(k) , b_c^2(k) \equiv \mathcal{P}_{cc}(k)/\mathcal{P}_{\rho\rho}(k) . \qquad (12)$$

The hope is that linear amplification holds over large scales, $i.e.$, that $b_g$ and $b_c$ are $k$-independent, and the power spectra inferred from redshift surveys reveal an underlying density spectrum [19]. A prediction is that the power in the cross-correlation of clusters and galaxies obeys $\mathcal{P}_{gc}(k) = b_g b_c \mathcal{P}_{\rho\rho}(k)$ [20]. Remarkably, the data are roughly consistent with this simple picture.

Another important unknown is the reheating history of the Universe, which may have a strong impact upon CMB anisotropies, and because it depends upon when and how efficiently massive stars formed in the pregalactic Universe, it is especially hard to predict in a given theory.

*1.2. The Observable Range in k-space*

For hierarchical theories of cosmic structure formation, we may roughly divide $k$-space into various wavebands shown in Fig.1.. (I normalize $a$ to be unity now so that comoving wavelengths, $2\pi k^{-1}$, are expressed in current cosmic length units. Since these are estimated from recession velocities, the unit is the h$^{-1}$Mpc. $a^{-1} - 1$ is the redshift at time $t$.) The astronomy associated with each band is: **ULSS** (ultra-large-scale-structure), with $k^{-1}$ in excess of a few times the Hubble radius, $cH_0^{-1} = 3000$ h$^{-1}$Mpc. We get mean Hubble patch values out of this ($i.e.$, "global" parameters such as $\Omega_{curv}$) and a little very long wavelength fluctuation information. **VLSS** (very-large-scale-structure), from the horizon scale ($k^{-1} \sim 2cH_0^{-1}$ for Einstein-deSitter models) down to say $k^{-1} \sim 100$ h$^{-1}$Mpc: $\delta\rho$ and $\delta v$ are apparently small enough that they exert little influence on observed cosmic structures, but gravitational potential perturbations generate large angle CMB anisotropies. **LSS** (large scale structure), from $\sim 100\,\mathrm{h^{-1}Mpc}$ down to about $5\,\mathrm{h^{-1}Mpc}$: we infer that the evolution of the waves in this band is sufficiently linear that first order perturbation calculations of the large scale streaming of galaxies and the clustering of galaxies and clusters may be valid, incomparably simpler than trying to correct for complex nonlinearities associated with dynamics and biasing. Fig.1. shows the wavebands probed by various large scale structure observations (large scale streaming velocities LSSV [21], the angular correlation of galaxies $w_{gg}(\theta)$, the power spectrum and redshift space correlation function of galaxies as probed by the QDOT and other redshift surveys *e.g.,* [22, 23, 25], the correlation function of clusters of galaxies $\xi_{cc}$ *e.g.,* [26, 27, 28]). The best indicator for large scale power is the angular correlation function of galaxies [29, 30].

CMB anisotropy experiments can be well characterized by filters which act upon a $k$-space 'power spectrum for $\Delta T/T$ fluctuations' [31]. Filter functions in $k$-space are shown for the COBE *dmr* ($\sim 7°$ beam) experiment [32], the *firs* (3.8° beam) balloon experiment [33], the UCSB *sp91* (1.5° beam) 'ACME-HEMT' South Pole experiments [34], the UCSB *sp89* (0.5° beam) experiment [35] and the Caltech OVRO *ov7* (1.8' beam) experiment [36]. The BigPlate Saskatchewan experiment [37] has a similar filter to *sp91*, the balloon-borne MAX [38, 39] and MSAM [40] experiments have filters which cover about the same range as *sp89*. WhiteDish [41] and a new OVRO (7' beam) experiment cover the region between *sp89* and *ov7*. Of more direct observational relevance are the corresponding filters in $\ell$-space shown in Fig.2.(b), showing experimental sensitivity to multipole components in the radiation anisotropy pattern.

Thus, CMB anisotropy experiments cover the entire VLSS and LSS bands. *Primary anisotropies* of the CMB are those one calculates from linear perturbation theory and which are therefore the most important ones because they are easiest to interpret. Their power spectra are quite complex [31, 3], because they include effects associated with geometrical ripples in the past light cone (Sachs-Wolfe effect), with the flow of electrons at photon decoupling, the degree of photon compression at decoupling, and the damping associated with the width of decoupling: below $\sim 5\,\mathrm{h^{-1}Mpc}$, the primary power is basically erased if hydrogen recombination is standard (SR line in Fig.1.); if there is an early injection of energy which ionizes the medium, photon decoupling would not have occurred until a lower ($\Omega_B$-dependent) redshift and would erase $\Delta T/T$ power on scales typically below the NR (no recombination) line shown.

Below **LSS** lie wavebands for which gas physics will have been extremely important, if not dominant, in determining the nature of the objects we see and how they are clustered. Fluctuations are nonlinear in these regimes. The light long-dashed filter curves at smaller scales show the bands probed by very small angle microwave background experiments, the VLA, the SCUBA array on the sub-mm telescope JCMT, and the OVRO mm-array. Although their beams are too small to see primary CMB anisotropies, they will provide invaluable probes of *secondary anisotropies* (those generated by nonlinear effects, including redshifted dust emission from galaxies and Thomson scattering from nonlinear structures in the pregalactic medium).

In a hierarchical model, nonlinearity at different scales will occur at sufficiently different epochs that I divide the "gastrophysical" realms into medium, small, very small and ultra small, bands ((**MSS, SSS, VSSS, USSS**), responsible for the construction of, respectively: clusters and groups ($\sim 10^{14-15}\,\mathrm{M_\odot}$); bright galaxies ($\sim 10^{11-12}\,\mathrm{M_\odot}$); dwarf galaxies and Lyman alpha clouds ($\sim 10^{9-10}\,\mathrm{M_\odot}$); and the first gas clouds to collapse ($\sim 10^{6-7}\,\mathrm{M_\odot}$), which make the first stars. Of course, significant gas dynamical processing may obscure the hierarchical relationship between object and primordial fluctuation waveband. Further, damping processes or tilted initial spectra may require some of the shorter distance structure to arise from fragmentation and other non-gravitational effects.

'Observed' power spectra (actually their square roots) are shown as hatched regions for density fluctuations inferred from COBE and for galaxy fluctuations inferred from the APM [29] and ROE [30] $w_{gg}$ data. The long wavelength hatched curve is the *dmr*-normalized scale invariant spectrum (assuming an $\Omega_{nr} = 1$ model, and including the current 10% *dmr* error on overall amplitude). The heavy curve extending the hatched $w_{gg}$ power into smaller distances is the power corresponding to the well known $\xi_{gg}(r) = (r/r_{0gg})^{-\gamma}$ 3D correlation function form, where the old CfA1 redshift survey values have been taken, $r_{0gg} = 5.4\,h^{-1}\rm Mpc$ and $\gamma = 1.8$. Power spectra derived from the QDOT [22], IRAS 1.2 Jansky [23] and CfA2 [25] redshift surveys are compatible with the range inferred from $w_{gg}$ when account is taken of redshift space distortions and biasing offsets between IRAS and optically identified galaxies. As already noted, cluster-cluster correlations and galaxy-cluster cross correlations [28] also seem to be compatible with this inferred spectrum.

The (linear) density fluctuation power spectra shown in Fig.1. are for three ($\Omega_{nr} = 1$) models normalized to the COBE *dmr* data : a standard CDM model with an initially scale-invariant spectra $n_s = 1$, one with the spectrum tilted to $n_s = 0.6$, and an $n_s = 1$ model whose shape is characterized by a parameter $\Gamma = 0.25$, whereas $\Gamma = 0.5$ for the standard CDM model. To fit the galaxy clustering data requires $0.15 \lesssim \Gamma \lesssim 0.3$ or $0.2 \lesssim n_s \lesssim 0.6$ (see § 5. for discussion). The biasing factor $b_g$ is relied upon to move the curves up into the allowed $w_{gg}$ band and nonlinearities to bend the shape upward to match the 1.8 law for $k^{-1} < 5\,h^{-1}\rm Mpc$. Although this LSS 'extra power' problem has been a subject of intense research on variations in the scale-invariant minimal-CDM theme for many years (*e.g.*, [20, 12]), we should bear in mind the great success inherent in the extrapolation over so many decades from COBE normalization to large and small scale structure formation: it seems scale invariance cannot be wildly broken and non-minimality cannot be too extreme, even if the generation mechanism has nothing to do with inflation (with the isocurvature baryon model being one deviant case [42]).

## 2. The Shape and Amplitude of the Primordial Post-Inflation Power Spectra

During inflation, the same zero point quantum fluctuation phenomenon which leads to the inflaton density perturbations also leads to statistically independent gravitational wave perturbations [43, 44]. The equations for the inflaton, isocons, and $\frac{m_{\mathcal{P}}}{\sqrt{16\pi}}h_{+,\times}$ are derived from identical scalar field actions, except the inflaton and isocons are coupled through a potential $V(\phi_{inf}, \phi_{is}, \ldots)$, while the gravity waves have no effective mass. Provided the effective masses of the scalars are small compared with $H^2$, all respond in basically the same way, rapid oscillation of the respective mode functions 'inside the horizon' ($k > Ha$), almost freeze-out outside ($k < Ha$), with a power amplitude $\mathcal{P}_\phi^{1/2}(k,t) \approx H/(2\pi)$ essentially given by the Hawking temperature on the $k = Ha$ boundary on time surfaces of uniform $(Ha)^{-1}$, a result motivated by a WKB treatment of fluctuation evolution inside the horizon. In stochastic inflation, noise at the Hawking temperature radiates from short distances across the decreasing $(Ha)^{-1}$ boundary into a long wavelength background field. We [12, 9] confirmed quantitatively that this simple picture, implicit in the early "new inflation" calculations of density perturbations, agrees with detailed numerical simulations of fluctuation generation. The inflaton fluctuations translate into scalar perturbations in the metric, codified in $\delta\ln a = (H/\Pi)\delta\phi$, where $\Pi$ is the background momentum of the inflaton field $\phi$. Thus, the post-inflation spectra are

$$\mathcal{P}_{GW}^{1/2} = \sqrt{8}\,\frac{\sqrt{4\pi}}{m_{\mathcal{P}}}\,\frac{H}{2\pi}\,e^{u_t}\,, \quad \mathcal{P}_{\ln a|_H}^{1/2} = \frac{1}{\sqrt{q+1}}\,\frac{\sqrt{4\pi}}{m_{\mathcal{P}}}\,\frac{H}{2\pi}\,e^{u_s}\,, \qquad (13)$$

The correction factors $u_t$ and $u_s$ to 'the $H/(2\pi)$ at $k = Ha$ WKB approximation' are in practice nearly zero. How near is now of considerable interest because the COBE results have created a desire for calculational precision (§ 2.1.).

$H(\phi)$ and $q(\phi)$ are treated as functions of the inflaton field here, which naturally follows from the Hamilton-Jacobi formulation of the problem [9, 45]: the solution to the momentum constraint equation, $\Pi = -\frac{m_{\mathcal{P}}^2}{4\pi}\left(\frac{\partial H}{\partial \phi}\right)$, is put into the energy constraint equation, turning it into the 'reduced Hamilton-Jacobi equation' relating $H(\phi)$ to the potential $V(\phi)$:

$$H^2 = \frac{H_{SR}^2}{1-(q+1)/3} \ , \ H_{SR}^2 \equiv \frac{8\pi V}{3m_{\mathcal{P}}^2} \ ; i.e., \ H^2 = \frac{8\pi}{3m_{\mathcal{P}}^2}\left[\frac{1}{2}\left(\frac{m_{\mathcal{P}}^2}{4\pi}\frac{\partial H}{\partial \phi}\right)^2 + V(\phi)\right] \ , \qquad (14)$$

with $H(\phi)$ taking the role of the (reduced) action. (Eq.(14) has corrections dependent upon the spatial curvature, hence is valid only for the smoothly varying (long wavelength) background field, not the sub-$(Ha)^{-1}$ fluctuating part.)

The power spectrum ratio and the adiabatic scalar and tensor indices follow:

$$(1+q) = \frac{n_t+3}{n_t+1} = \frac{m_{\mathcal{P}}^2}{4\pi}\left[\frac{\partial \ln H}{\partial \phi}\right]^2 = e^{-(u_t-u_s)}\frac{1}{8}\frac{\mathcal{P}_{GW}}{\mathcal{P}_{\ln a|_H}} \ , \qquad (15)$$

$$\frac{n_t+3}{2} = 1+q^{-1}+C_t \ , \qquad (16)$$

$$\frac{n_s-1}{2} = 1+q^{-1}-q^{-1}\frac{m_{\mathcal{P}}^2}{4\pi}\frac{\partial^2 \ln H}{\partial \phi^2} + C_s = 1+q^{-1}-q^{-1}\frac{\text{sgn}(\partial H/\partial \phi)}{\sqrt{1+q}}\frac{m_{\mathcal{P}}}{\sqrt{4\pi}}\frac{dq}{d\phi}+C_s \ .$$

The accurate path to the spectral indices is to take logarithmic derivatives of full numerical calculations, *a la* [12]. The stochastic inflation technique [9] is to write eq.(13) as a function of $H$, $q$ and derivatives, and take a logarithmic derivative *wrt $Ha$* in place of $k$, the path adopted here and in [9, 2, 3, 5]. Eq.(16) shows that tilt mostly depends upon how far the acceleration is below the *critical* value of unity (but for $q \approx -1$, a substantial scalar tilt can come from the second term, yet no tensor tilt, as in § 2.4.). Here $C_{t,s}$ are correction factors associated with derivatives of the $u_{t,s}$, which I now discuss in § 2.1., but which the reader may wish to skip since I find them to be small and thus drop them subsequently.

*2.1. Corrections to the Stochastic Inflation Calculation of Power Spectra and Their Shapes*

We and others have often used the $u_{t,s} = 0, C_{t,s} = 0$ approximation *e.g.*,, [2, 3, 5], but [44, 15, 47, 48, 46, 49] have stressed the importance of higher order corrections. For the case of **uniform acceleration**, the tensor and scalar equations can be solved analytically in terms of Hankel functions and the asymptotic limit can be taken to determine the correction factors for the tensor [44] and scalar [15] modes:

$$u_t = \ln \Gamma(\tfrac{1}{2}-q^{-1}) - q^{-1}\ln(-2q) - \tfrac{1}{2}\ln \pi \ , \qquad (17)$$

$$\to u_t'(1)(1+q) + \mathcal{O}(1+q)^2 \ ,$$

$$\approx u_t'(\infty) - \nu(1-\ln 2) + \tfrac{1}{2} - \frac{\nu^{-1}(1+\nu^{-1}/2\ldots)}{24} \ \text{for large } \nu \equiv (\tfrac{1}{2}-q^{-1}) \ ,$$

$$u_t'(-q^{-1}) \equiv \frac{du_t}{d(-q^{-1})} = -(\gamma+\ln 2-1) + \left[\Psi(\tfrac{1}{2}-q^{-1}) - \Psi(\tfrac{3}{2}) - \ln(-q^{-1})\right] \qquad (18)$$

$$\approx -(1-\ln 2) + \frac{\nu^{-2}(1+\nu^{-1}+\ldots)}{24} \ , \ \text{large } \nu \ .$$

Here $\Psi(\nu) = d\ln\Gamma/d\nu$ is the diGamma function, $2-\gamma-2\ln 2 = 0.03649$ at $3/2$, where $\gamma$ is Euler's constant. The large $\nu$ limit is surprisingly useful: at $-q^{-1} = 1$, it is only off by 2% and quickly gets better. There are also weak corrections associated with acceleration changes and the effective masses of the scalars, which [46] dealt with by assuming slow changes of these quantities to exploit the uniform acceleration analytic solution, not strictly valid but useful to indicate the correction level. Following this path, and keeping the leading $dq/d\phi$ term, which is the only important one, we have

$$u_s \approx u_t + (u_t'(-q^{-1})+1)\frac{m_{\mathcal{P}}^2}{4\pi}\frac{\partial^2 \ln H}{\partial \phi^2} \ . \qquad (19)$$

Thus the correction factors are

$$C_t = 2q^{-2}u'_t(-q^{-1})\,(1+q)\,q^{-1}\frac{m_{\mathcal{P}}^2}{4\pi}\frac{\partial^2 \ln H}{\partial\phi^2}\,, \tag{20}$$

$$C_s = C_t + \text{sgn}(\partial H/\partial\phi)\,(u'_t(-q^{-1})+1)\,(1+q)^{\frac{1}{2}}q^{-2}\frac{m_{\mathcal{P}}^3}{(4\pi)^{3/2}}\frac{\partial^3 \ln H}{\partial\phi^3}\,. \tag{21}$$

For evaluations, substituting $-0.3$ for $u'_t(-q^{-1})$ provides enough accuracy ($u'_t(1) = -0.27$, $u'_t(\infty) = -0.31$). The key point in $C_t$ is the $(1+q)$ multiplier, which effectively suppresses this term relative to the $\partial^2 \ln H/\partial\phi^2$ term of eq.(16): the ratio is $2q^{-2}u'_t(-q^{-1})\,(1+q) \approx -0.6(1+q)$. In § 5.2., we find the data suggests we restrict our attention to tilts $\lesssim 0.2$, hence this ratio is below 7%. And when the $\partial^2 \ln H/\partial\phi^2$ is most important in eq.(16) is when $q \approx -1$, as for natural inflation, § 2.4., and in this case, the $C_t$ correction is exponentially suppressed. The $\partial^3 \ln H/\partial\phi^3$ correction to $n_s$ has a less strong suppression factor, $(1+q)^{1/2}$, but effective enough. An advantage of the forms adopted here over those in [49] is that one is not restricted to the $(1+q) \approx 0$ regime. But, for the reasons given, I believe it is safe to drop them, which I now do.

### 2.2. Uniform Acceleration: Exponential Potentials and Extended Inflation

A constant acceleration regime implies equal scalar and tensor tilts and power law inflation ($a \propto t^p$):

$$q+1 = p^{-1}\,,\quad n_s - 1 = n_t + 3 = -2(p-1)^{-1}\,. \tag{22}$$

Eqs.(14,15) implies an exponential potential, $V = V_0 \exp[-\sqrt{16\pi(q+1)}\,\phi/m_{\mathcal{P}}]$. Of course, $q$ must go negative for a viable model of inflation. Nonetheless, over the observable $k$-range, the exponential approximation is often quite good, even when rather drastic potential surfaces are adopted to 'design' spectra.

Theories with $f(\phi)R$ couplings, where $R$ is the curvature, and with one or more dynamically important scalar fields are a rich source of inflation models. The classical Brans-Dicke theory has $f = \phi^2/(4\omega)$, where $\phi$ is related to the dilaton. In [12], we considered the induced gravity model, with $\phi$ as the inflaton, and showed that if $\omega \sim 10^{-5}$ in the early universe, the coupling of all fields would be effectively weak and the observed density fluctuation level would result. However an arbitrary symmetry breaking potential was invoked to eventually pin $\phi$ at $\phi = m_{\mathcal{P}}/\sqrt{4\pi}$ to get the observed Newton gravitational 'constant'. In extended inflation [50], the inflaton is a separate degree of freedom from the dilaton. The deceleration in the conformally transformed frame is uniform, e.g., [51], with value $q+1 = 4/(2\omega+3)$, hence $n_s - 1 = n_t + 3 = 8/(2\omega-1)$. Thus another mechanism was required for reheating, bubble nucleation. However, to avoid an excessive CMB anisotropy due to large bubbles, the theory needed $\omega \lesssim 18-25$ at the end of inflation, yet to satisfy solar system tests $\omega \gtrsim 500$ and hence an effective $\omega$-pinning or $m_{\mathcal{P}}$-pinning mechanism is required.

With conformal transformations, the kinetic term can become nontrivial, making the standard Hamilton-Jacobi derivation leading to eq.(15), which predicts $q \geq -1$, incorrect. We explore the implications of a $q < -1$ supercritical acceleration for the shape elsewhere [52], but for this paper I shall consider only cases for which field reparameterization can take the kinetic piece into the standard form, and for which eq.(15) is valid.

### 2.3. Slowly Dropping Acceleration: Chaotic Inflation and Power law Potentials

Power law potentials of the form $V(\phi) = \lambda_e m_{\mathcal{P}}^4(\phi/m_{\mathcal{P}})^{2\nu}/(2\nu)$ have the advantage over exponential laws that $q \approx -1 + (\phi/m_{\mathcal{P}})^{-2}\nu^2/(4\pi)$ naturally passes through zero. Chaotic inflation discussions [53] have typically focussed on simple potentials, in particular the power law form with $\nu$ taken to be 1 or 2. A characteristic of such potentials is that the range of values of $\phi$ which correspond to all of the large scale structure that we observe is actually remarkably small: e.g., for $\nu = 2$, the region of the potential curve responsible for the structure between the scale of galaxies and the scales up to our current Hubble length is just $4m_\mathcal{P} \lesssim \phi \lesssim 4.4m_\mathcal{P}$ [12]. Consequently, $H(\phi)$ does not evolve by a large factor over the large scale structure region and we therefore expect approximate uniformity of $n_s(k)$ and $n_t(k)$ over the narrow observable bands of $k$-space, and near-scale-invariance for both. Although this is usually quoted in the form of a logarithmic correction to the $\ln a|_H$-spectrum, a power law approximation is quite accurate [2]:

$$q + 1 \approx \frac{\nu/2}{N_I(k) + \nu/3} \;,\; n_s(k) \approx 1 - \frac{\nu+1}{N_I(k) - \nu/6} \;,\; n_t \approx -3 - \frac{\nu}{N_I(k) - \nu/6} \;. \quad (23)$$

$N_I(k)$ is the number of e-foldings from the point at which wavenumber $k$ 'crosses the horizon' (when $k = Ha$) and the end of inflation. For waves the size of our current Hubble length we have the familiar $N_I(k) \sim 60$, hence $n_s \approx 0.95, n_t \approx 0.97$ for $\nu = 2$ and $n_s \approx 0.97, n_t \approx 0.98$ for $\nu = 1$ (massive scalar field case). Further, the observable scales are sufficiently far from the reheating scale that $N_I$ is relatively large over the observable range: e.g., over the range from our Hubble radius down to the galaxy scale, $n_s$ decreases by only about 0.01.

### 2.4. Dropping from Nearly-Critical Acceleration: Natural Inflation

In natural inflation [54, 2], the inflaton for the region of $k$-space that we can observe is identified with a pseudo-Goldstone boson with a potential $V = 2\Lambda^4 \sin^2(\phi/(2f))$. This is similar to the axion, except that the symmetry breaking scale $f$ is taken to be of order $m_\mathcal{P}$ and the energy scale for the potential is taken to be of order the grand unified scale, $m_{GUT}$, so that an effective weak coupling, $\lambda_e = \Lambda^4/(fm_\mathcal{P})^2 \sim (m_{GUT}/m_\mathcal{P})^4$ arises 'naturally', giving the required $10^{-13}$ for $m_{GUT} = 10^{16}$GeV. To obtain sufficient inflation and a high enough post-inflation reheat temperature for baryogenesis, $f \gtrsim 0.3m_\mathcal{P}$ is required.

To have a tilted spectrum and also get enough inflation in our Hubble patch, $\phi/f$ must have started near the maximum at $\pi$, an inflection point where $q$ is nearly $-1$ [2]:

$$q + 1 \approx (1 + \frac{m_\mathcal{P}^2}{24\pi f^2}) \exp\big[-\frac{m_\mathcal{P}^2}{8\pi f^2} N_I(k)\big] \approx 0 \;, \qquad (24)$$

$$n_s(k) \approx 1 - \frac{m_\mathcal{P}^2}{(8\pi f^2)} + \mathcal{O}(q+1) \;,\; n_t + 3 \approx \mathcal{O}(q+1) \;,\; \frac{\mathcal{P}_{GW}}{\mathcal{P}_{\ln a|_H}} \approx \mathcal{O}(q+1) \;. \quad (25)$$

Thus, we can have a scalar tilt but tensor tilt and gravity wave power are both exponentially-suppressed.

### 2.5. Rapid Acceleration Changes: Radically-Broken Scale Invariance

The index can have complex $k$-dependent structure when the acceleration changes considerably over the $k$-band in question. According to eqs.(16), the post-inflation gravitational wave spectrum will have power increasing with wavelength (the correction $C_t$ seems unlikely to modify this, although supercritical acceleration can), whereas artfully using the $\partial\sqrt{1+q}/\partial\phi$ term in the inflaton effective potential allows essentially any prescribed shape for the adiabatic scalar spectrum.

*A priori*, it seems unlikely that a marked change in the expansion rate or acceleration would just happen to be in the narrow window of $k$-space accessible to our observations. However, in $\phi$-space, this window is not very far from $\phi_{end}$ defining the acceleration/deceleration boundary, hence the $q$ rise to zero must be reasonably rapid in $\phi$. Even so, for the models described above, the rapid change does indeed occur only near the end, suggesting special physics might have to be built in.

Rapid acceleration changes, if present, would seem to be more likely a consequence of interaction with other field degrees of freedom rather than a result of inflaton self-interaction. Thus, many of the toy models constructed to illustrate that radically broken scale invariance is possible involved two scalars interacting with either simple polynomial potentials (with second, third and fourth order terms) [55, 12], or combinations of exponential potentials [47].

Even with many scalar fields being dynamically important, it is often possible to consider an effective single inflaton self-interacting through an effective single-inflaton potential over the observable scales. This is because the fields first settle into gorges on the potential surface, then follow the gorge downward towards the local minimum along a single field degree of freedom, $\phi_\parallel$, to be identified with the inflaton. The other degrees of freedom, $\vec{\phi}_\perp$, are 'isocurvature' degrees of freedom. Usually, the faces rising up from the gorge will be sufficiently steep that the inevitable quantum noise that excites motion up the walls quickly falls back, leaving no usable isocurvature imprint, effectively making those dimensions irrelevant (although curvature in the trough can lead to complications in the kinetic energy piece of the inflaton degree of freedom).

Many models of double inflation could be described this way, consisting of two periods of inflation with relatively constant $H$, one at high $H$, the other at low $H$. These lead to nearly scale invariant fluctuations in the two associated regions of $k$-space, high amplitude, then low. The join *must* be accompanied by a large change in acceleration, hence in $n_{s,t}(k)$ over the corresponding $k$-band: exactly how one crafts the transition determines the detailed form of $n_{s,t}(k)$. General variations of the effective single inflaton potential $H(\phi)$, hence of $V(\phi)$, allow wide latitude in what can be constructed. Since $n_s$ has a term $\propto \partial\sqrt{1+q}/\partial\phi$, it tends to be more susceptible to the variations than $n_t$ is, and therefore the adiabatic scalar spectra should exhibit sharper structural features.

Models with two scalar fields that do not allow an effective single inflaton approximation over the relevant band in $k$-space have also been used to construct power spectra with mountains and valleys and also to generate non-Gaussian inflation fluctuations. Often these involve an instability, with negative transverse components of the mass-squared matrix, $\partial^2 V/\partial\phi_i\partial\phi_j$, leading to an opening up of the gorge or its bifurcation. Tuning the location of such a structure to the window on the potential surface we can access must be unpalatably precise [12].

## 3. Inflation-Based CMB Power Spectra c.f. the Data

### 3.1. Theoretical CMB Power Spectra

For a given inflation model, perturbed Einstein-Boltzmann equations (*e.g.*, [56, 31, 3]) must be solved for each mode $\mathcal{M}$ present to get the temperature radiation pattern at our location and at this time:

$$\frac{(\Delta T)^{(\mathcal{M})}}{T_0}(\hat{q}, \text{here}, \text{now}) \equiv \frac{\delta f_\gamma^{(\mathcal{M})}}{T_\gamma \partial \bar{f}_\gamma/\partial T_\gamma} = \sum_{\ell m} a_{\ell m}^{(\mathcal{M})} Y_{\ell m}(\hat{q}) \,, \tag{26}$$

where $\bar{f}_\gamma$ is the unperturbed Planck distribution and $\delta f_\gamma^{(\mathcal{M})}$ is the distribution function fluctuation. If the post-inflation fluctuations are Gaussian-distributed, then so are the multipole coefficients $a_{\ell m}^{(\mathcal{M})}$, with amplitudes fully determined by just the angular power spectra $\mathcal{C}_\ell^{(\mathcal{M})}$, defined by

$$\mathcal{C}_\ell^{(\mathcal{M})} \equiv \frac{\ell(\ell+1)}{2\pi} \langle |a_{\ell m}^{(\mathcal{M})}|^2 \rangle \,. \tag{27}$$

Sample theoretical $\mathcal{C}_\ell$'s are shown in Fig.2.(a). The "standard" adiabatic CDM model ($\Omega = 1$, $n_s = 1$, h = 0.5, $\Omega_B = 0.05$) with normal recombination illustrates the typical form: the Sachs-Wolfe effect dominating at low $\ell$, followed by rises and falls in the first and subsequent Doppler peaks, with an overall decline due to destructive interference across the photon decoupling surface. A similar CDM model, but with early reionization (at $z > 200$), shows the Doppler peaks are damped, a result of destructive interference from forward and backward flows across the decoupling region, illustrating that the "short-wavelength" part of the density power spectrum can have a dramatic effect upon $\mathcal{C}_\ell$, since it determines how copious UV production from early stars was. Lower redshifts of reionization still maintain a Doppler peak, but suppressed relative to the standard CDM case.

A form often adopted to describe the low-$\ell$ end is the "Sachs-Wolfe" power for scalar metric perturbations [31]

$$\begin{align}
\mathcal{C}_\ell &= \frac{1}{9} \int d\ln k\, \mathcal{P}_\Phi(k) j_\ell^2(kR_r) , \tag{28} \\
&\approx \frac{1}{9} \mathcal{P}_\Phi(R_n^{-1}) \left(\frac{R_r}{R_n}\right)^{1-n_s} \frac{\Gamma(\ell - \frac{1-n_s}{2})\Gamma(\ell+2)}{\Gamma(\ell)\Gamma(\ell+2+\frac{1-n_s}{2})} \frac{\Gamma(1-\frac{1-n_s}{2})\Gamma(\frac{3}{2})}{\Gamma(\frac{3}{2}+\frac{1-n_s}{2})} , \tag{29} \\
&\approx \frac{1}{9} \mathcal{P}_\Phi(R_n^{-1}) \left(\frac{R_r}{R_n}\right)^{1-n_s} (\ell+\tfrac{1}{2})^{n_s-1} \frac{\Gamma(1-\frac{1-n_s}{2})\Gamma(\frac{3}{2})}{\Gamma(\frac{3}{2}+\frac{1-n_s}{2})} \left(1 + \frac{\varepsilon}{(\ell+\frac{1}{2})^2} + \ldots\right), \\
\varepsilon &= \frac{(1-n_s)(11 + 6(1-n_s) + (1-n_s)^2)}{24} .
\end{align}$$

Here $R_r$ is the comoving distance from us to the surface over which photons decouple from the baryons, about $5800\,h^{-1}$Mpc away for a CDM cosmology, and $R_n$ is a normalization scale. The analytic result in terms of Gamma functions holds if there are no deviations from the power law form for the power spectrum of the gravitational potential $\mathcal{P}_\Phi$, related to $\mathcal{P}_{\ln a|H}$ by $\mathcal{P}_{\ln a|H}^{1/2} \approx (5/3)\mathcal{P}_\Phi^{1/2}$. This predicts a flat $\mathcal{C}_\ell$ for $n_s = 1$, obeying the pleasing formula $\mathcal{C}_\ell^{1/2} = \mathcal{P}_\Phi^{1/2}/3 \approx \mathcal{P}_{\ln a|H}^{1/2}/5$, relating $\Delta T/T$ to $\Phi/3$. In fact, for a realistic model, there are corrections to this formula from other anisotropy sources: in particular for the standard CDM model shown, there is a small rise over the multipoles that COBE probes, modelled by an effective index 1.15 if we use the eq.(29) form.

The dotted $\mathcal{C}_\ell$ in Fig.2.(a) also has a flat initial spectrum, but has a large nonzero cosmological constant in order to have a high $H_0$, in better accord with most observational determinations. The specific model has $\Omega_\Lambda = 0.75, \Omega_{cdm} = 0.22, \Omega_B = 0.03, H_0 = 75, n_s = 1$. As one goes from $\ell = 2$ to $\ell = 3$ and above there is first a drop in $\mathcal{C}_\ell$ [57], a consequence of the time dependence of $\Phi$ which results in corrections to $\mathcal{C}_\ell^{1/2} \sim \mathcal{P}_\Phi^{1/2}/3$.

A major goal of CMB anisotropy research is to determine all of the ups and downs of the spectrum in detail. The 'data points' of Fig.2.(a) are averages of the $\mathcal{C}_\ell$'s *wrt* 'filter' functions $W_\ell$:

$$\langle \mathcal{C}_\ell \rangle_W \equiv \mathcal{I}[\mathcal{C}_\ell W_\ell]/\mathcal{I}[W_\ell] , \text{ where } \mathcal{I}[f_\ell] \equiv \sum_\ell \frac{(\ell+\tfrac{1}{2})}{\ell(\ell+1)} f_\ell \tag{30}$$

defines the "logarithmic integral" $\mathcal{I}[f_\ell]$ of a function $f_\ell$. The $W_\ell$ are taken to be those for a set of existing anisotropy experiments spanning the range $10°$ to $2'$ shown in Fig.2.(b): $\langle \mathcal{C}_\ell \rangle_W$ then characterizes the broad-band power that the experiment is sensitive to. The location in $\ell$-space is $\langle \ell \rangle_W$. The error bars are 10%, the best *dmr* can possibly do with the full 4 years of data, and ones that are actually quite reasonable for intermediate- and small-angle mapping experiments. The observed data points for these experiments are shown in Fig.4. below.

If the spectrum is tilted, gravitational waves will generally be present to induce a tensor-mode spectrum, $\mathcal{C}_\ell^{(t)}$ to add to the adiabatic scalar spectrum $\mathcal{C}_\ell^{(s)}$. The amplitude of gravitational wave modes decays by directional dispersion as the modes re-enter the horizon, just as waves in any relativistic collisionless matter do. Before the gravitational waves disperse however, they influence the microwave background through the inhomogeneous redshifts they induce. The recognition of the potential importance of gravity waves for large angle microwave background fluctuations has a long history, e.g., [43, 44] and generated a tremendous post-COBE burst of excitement, and papers e.g., [61, 47, 48, 62, 2], especially when the sensitivity to tilt was realized that meant the *dmr*-signal could even be largely tensor-induced. I shall parameterize the relative magnitudes of scalar and tensor by the ratio of the broad-band powers associated with the *dmr*-beam's filter:

$$\tilde{r}_{ts} \equiv \frac{\langle \mathcal{C}_\ell^{(t)} \rangle_{dmr}}{\langle \mathcal{C}_\ell^{(s)} \rangle_{dmr}} \approx 1.5 \mathcal{P}_{GW}/\mathcal{P}_{\ln a|H} \approx 6 \frac{n_t + 3}{(n_t + 1)/2} \ . \quad (31)$$

This ratio has no simple analytic result and its value is dependent upon the details of the cosmology being considered. The 1.5 numerical result holds for small deviations from scale invariance, $n_s \approx n_t$, and for CDM-like models. (In [3, 5], we used instead $\mathcal{C}_2^{(t)}/\mathcal{C}_2^{(s)} \approx 1.2 \tilde{r}_{ts}$ to characterize the relative magnitudes.)

In Fig.2.(a), the dashed line shows a standard CDM model, but with a chaotic-inflation inspired $n_s = 0.95$ tilt, along with a $\tilde{r}_{ts} \approx 0.3$ gravity wave contribution (eq.(31) with $n_t = n_s$), contributing the lower dashed line to the total [3]. Although $\mathcal{C}_\ell^{(t)}$ has a flat part at low $\ell$ just as $\mathcal{C}_\ell^{(s)}$ does for this nearly scale invariant model, there is about a 20% drop from $\ell = 2$ to $\ell = 3$, and there is no Doppler peak, only a rapid decline at $\ell \gtrsim 50$.

To get a visual impression of what the spectral structure means, Fig.3. shows what the sky looks like on a few resolution scales for the standard $n_s = 1$ CDM model: on the COBE beamscale (Gaussian filtering $\ell_s = 19$), the nearly scale invariant form; on the half-degree scale ($\ell_s = 269$), where the standard recombination spectrum is a maximum; with no smoothing at all, with the shapes defined entirely by the destructive interference that occurred across the photon decoupling region. For early-reionization, the shapes in the 60° NR map are also the naturally occurring ones, since there is no power left at $\ell_s \sim 269$ to artificially filter.

*3.2. How Accurately Can The Spectra Be Measured?*

In [5], we showed that for small variations about the CDM model, the height of the first Doppler peak relative to the *dmr* band-power is (within $\sim 15\%$)

$$\frac{\mathcal{C}_\ell|_{max}}{\langle \mathcal{C}_\ell \rangle_{dmr}} \approx 5 \, e^{-3.6(1-\tilde{n}_s)} \, , \ \tilde{n}_s \approx n_s - \ln(1+\tilde{r}_{ts})/3.6 - 0.5[(1-\Omega_\Lambda)^{\frac{1}{2}} h - \tfrac{1}{2}] - \left(\frac{1+z_R}{200}\right)^{3/2} \, , \quad (32)$$

where $\Omega_B = 0.0125 h^{-2}$ has been fixed at the standard BBN value. Here $z_R$ is the reionization redshift and must be $\lesssim 150$ to have a local maximum. For example, a model with no gravity wave contribution (as natural inflation would predict) but $n_s \approx 0.88$ has a spectrum that is almost degenerate with the $n_s = 0.95$, $\tilde{r}_{ts} = 0.3$ spectrum, so much so that it will be extremely difficult to tell them apart. More generally, we argued that the precision required to separately determine $n_s, \tilde{r}_{ts}, \Omega_\Lambda, \ldots$ is too high for likely experiments, but $\tilde{n}_s$ can determined accurately. (An exception is $\Omega_B h^2$, which the relative heights of the Doppler peaks are sensitive to.) To separate the various contributions to $\tilde{n}_s$ requires other cosmological experiments, e.g., measuring the scalar perturbation shape through galaxy-galaxy power spectra (§ 5.1.) and amplitude through cluster abundances or streaming velocities (§ 5.2.); and, in some happy future, determining $H_0$ definitively.

We now discuss the experiments in more detail, first in an idealized way to show what is needed for achieving even the 10% error bars shown on the band-powers in Fig.2.(a). The signal $(\Delta T/T)_p$ from the $p^{th}$ pixel of a CMB anisotropy experiment can be expressed in terms of linear filters $\mathcal{F}_{p,\ell m}$ acting on the $a_{\ell m}$: $(\Delta T/T)_p = \sum_{lm} \mathcal{F}_{p,\ell m} a_{\ell m}$. The $\mathcal{F}_{p,\ell m}$ encodes the experimental beam and the switching strategy that defines the temperature difference, the former filtering high $\ell$, the latter low $\ell$. The pixel-pixel correlation function of the temperature differences can be expressed in terms of a quadratic $N_{pix} \times N_{pix}$ filter matrix $W_{pp',\ell}$ acting on

$C_{T\ell}$:

$$C_{Tpp'} \equiv \langle\left(\frac{\Delta T}{T}\right)_p \left(\frac{\Delta T}{T}\right)_{p'}\rangle = \mathcal{I}\left[W_{pp',\ell}\mathcal{C}_{T\ell}\right] , \quad W_{pp',\ell} \equiv \frac{4\pi}{2\ell+1}\sum_{lm}\mathcal{F}_{p,\ell m}\mathcal{F}^*_{p',\ell m} ; \quad (33)$$

$$\sigma^2_T[\overline{W}] \equiv \left(\frac{\Delta T}{T}\right)^2_{rms} \equiv \frac{1}{N_{pix}}\sum_{p=1}^{N_{pix}} C_{Tpp} \equiv \mathcal{I}\left[\overline{W}_\ell \mathcal{C}_{T\ell}\right] , \quad \overline{W}_\ell \equiv \frac{1}{N_{pix}}\sum_{p=1}^{N_{pix}} W_{pp,\ell} . \quad (34)$$

The trace $\overline{W}_\ell$ defines the average filters [58, 59, 60] shown in Fig.2.(b), which determine the *rms* anisotropies $\sigma_T[\overline{W}]$. Typically we will be given the anisotropy data in the form $\overline{(\Delta T/T)}_p \pm \sigma_{Dp}$, where $\sigma_{Dp}$ is the variance about the mean for the measurements. In general, there may be pixel-pixel correlations in the noise, defining a correlation matrix $C_{Dpp'}$ with off-diagonal components as well as the diagonal $\sigma^2_{Dp}$.

In the simplest experiment that can be imagined, we would have $C_{Dpp'} = \sigma^2_D \delta_{pp'}$ and the pixels sufficiently separated on the sky that only $\overline{W}_\ell$ is an effective probe of $\mathcal{C}_\ell$; i.e., that $C_{Tpp'} \approx \sigma^2_T \delta_{pp'}$. $\langle\mathcal{C}_\ell\rangle_{B,th} \equiv \sigma^2_T/\mathcal{I}(\overline{W}_\ell)$ is the quantity we wish to estimate. For large $N_{pix}$, the 1-sigma uncertainty in the experimental value of the band-power is [5, 6, 7]

$$\langle\mathcal{C}_\ell\rangle_{B,obs} = \langle\mathcal{C}_\ell\rangle_{B,maxL} \pm \sqrt{2/N_{pix}}\left[\langle\mathcal{C}_\ell\rangle_{B,maxL} + \sigma^2_D/\mathcal{I}[\overline{W}_\ell]\right] ; \quad (35)$$

$$\langle\mathcal{C}_\ell\rangle_{B,maxL} = \langle\mathcal{C}_\ell\rangle_{B,th} \pm \sqrt{1/N_{pix}}\left[\langle\mathcal{C}_\ell\rangle_{B,th} + \sigma^2_D/\mathcal{I}[\overline{W}_\ell]\right] . \quad (36)$$

An experimental noise $\sigma_D$ below $10^{-5}$ is standard now, and a few times $10^{-6}$ is soon achievable; hence, if systematic errors and unwanted signals can be eliminated, the 1-sigma relative uncertainty in $\langle\mathcal{C}_\ell\rangle_B$ will be from cosmic-variance alone, $\sqrt{2/N_D}$, falling below 10% for $N_{pix} = 200$, i.e., a mapping experiment. For large $N_{pix}$, the observed maximum likelihood will fluctuate from $\langle\mathcal{C}_\ell\rangle_{B,th}$, the quantity we want, according to eq.(36), but the error bars of eq.(35) include these realization-to-realization fluctuations (thus $\sqrt{2}$ appears, not 1). If there were full-sky coverage and errors from cosmic variance alone, the fractional error in the $\langle\mathcal{C}_\ell\rangle_W$ goes as $\sim \langle\ell\rangle^{-1}$ [7]: so tiny for intermediate and small angle experiments that it would appear that even extremely subtle differences in the spectra could in principle be determined at high $\ell$. Now I shall discuss how we are doing so far in practice.

### 3.2.1. *Current Status: Experimental Broad-Band Powers*

To determine band-powers for an experiment [4, 7], I construct a local model of $\mathcal{C}_\ell$, assumed to be valid over the scale of the experiment's average filter $\overline{W}_\ell$:

$$\mathcal{C}_{B\ell} = \langle\mathcal{C}_\ell\rangle_B (\ell+\tfrac{1}{2})^{2+n_{\Delta T}} \quad \mathcal{I}[\overline{W}_\ell]/\mathcal{I}[\overline{W}_\ell(\ell+\tfrac{1}{2})^{2+n_{\Delta T}}] , \quad n_{\Delta T} \approx n_s - 3 . \quad (37)$$

The local "spectral colour index", $n_{\Delta T}$, is similar to the $n_D$ of eq.(4), except for 2D, not 3D. The form differs *very* little from that in eq.(29): for small $\ell$, $n_{\Delta T}$ is related to $n_s$ as shown, but with $n_s$ now interpreted as a phenomenological rather than a primordial index. Thus $n_s = 3$ corresponds to white noise in $\Delta T$. To get the band-power, I use eq.(37) with $n_{\Delta T} = -2$, i.e., scale-invariant over the band, but check that the result is robust to variations in $n_{\Delta T}$. This is true for all intermediate and small angle experiments to date, and as we shall see, even holds true for *dmr* and *firs*, which have such a large coverage in $\ell$-space that they can also be used to determine the index $n_{\Delta T}$. The amplitude $\langle\mathcal{C}_\ell\rangle_B^{1/2}$ can be determined by whatever statistical method we are most enamoured with, whether Bayesian as I prefer, or frequentist.

There are so many detections now that I split Fig.4. into two panels for clarity, the upper giving the overview, the lower focusing on the crucial first Doppler peak region. These figures have been evolving rapidly since I introduced them [4, 7]. Data points either denote the maximum likelihood values for the band-power and the error bars give the 16% and 84%

Bayesian probability values (corresponding to $\pm 1\sigma$ if the probability distributions were Gaussian) or are my translations of the averages and errors given by the experimental groups to this language. Upper and lower triangles denote 95% confidence limits unless otherwise stated. See [7] for details. The horizontal location is at $\langle \ell \rangle_W$ and the horizontal error bars denote where the filters have fallen to $e^{-0.5}$ of the maximum (with Fig.2.(b) providing a more complete representation of sensitivity as a function of $\ell$). Only wavelength-independent Gaussian anisotropies in $\Delta T/T$ are assumed to be contributing to the signals, but non-primary sources (*e.g.*, dust, synchrotron) may contribute to these $\mathcal{C}_\ell$'s (as can unknown systematic errors of course). Cleaning $\Delta T/T$ observations has been done to some extent in most of these cases, and will be the key to the ultimate accuracy that we can achieve in spectrum determination - not theoretical cosmic variance. Generally the underlying primary $\mathcal{C}_\ell$ will be lower than the values shown, but it can be higher because of 'destructive interference' among component signals.

Proceeding from small $\ell$, the $\ell = 2$ power uses the first year 53GHz quadrupole values with a Galactic cut $b_{Gcut} = 20°$ [4]. The value for the combined first and second year data will be even lower [63], but it is also the multipole most likely to have a residual Galactic signal contaminating it, possibly destructively. The solid *dmr* point with the tiny error bar is my translation of the combined first and second year result [63], while the open *dmr* point for the first year data and the *firs* point are my band-powers, from refs.[4, 7] and § 4.. The Tenerife point uses the data for the limited region of the sky they probed at both 15 and 33 GHz. Remarkably, in view of the relatively low frequency, the band-power for their data at 15 GHz only, which covered a much larger region of the sky, agrees. We now come to the confused region from two degrees to half a degree, which can be better seen in the lower panel. Two *sp91* band-powers are shown, for a 9 point scan and a 13 point scan [34]. All 4 channels were simultaneously analyzed [4]. The offsets are for clarity. The BigPlate result [37], *bp*, is slightly higher. New *sp94* results apparently also have detections at a little higher level and there will also shortly be a new *bp* result, both with more extended frequency coverage. Python [65], *py*, has wide coverage in $\ell$-space, but has only the single 90 GHz frequency so signal cleaning cannot be done. Argo [66], *ar*, is next. The next 3 results are for 3 scans from the fourth flight of the MAX balloon-borne experiment [39], *M4*, the open squares for the Iota Draconis scan and the Sigma Hercules scan, the solid point for a GUM scan. Because the filters changed with frequency, the points are placed at the average over all MAX4 filters. The two solid data points are for the third MAX flight [38], the upper for a GUM scan, the lower for a Mu Pegasus scan (which had a strong dust signal removed). The GUM point also shows the Bayesian $1-sigma$ error bar. The dotted lines ending in triangles denote the 90% limits for the MSAM [40] single (*g2*) and double (*g3*) difference configurations for all of their data, although there are some worries that half of their data was contaminated by non-Gaussian sources, which, when excluded, lowers the band-powers somewhat. The next three give upper limits, but no detection. The open triangle is the 95% credible limit for the *sp89* 9 point scan [35, 60]. Switching back to the upper panel, the 95% limit from the $m=2$ mode analysis of the WhiteDish experiment [41] is *wd2*. There is a hint of a detection in the $m=1$ mode analysis. The *ovro* 7 point upper limit [36] is last. New *ovro* experiments with significantly higher sensitivity for this filter and for the *ovro22* one in Fig.2.(b) also have results that will soon be available, once the detections have been cleaned of radio sources.

In the future we will be able to strongly select the preferred theories by simultaneously analyzing experiments like these. Although we have already tried this, *e.g.*, for *sp89* and *ov7* [60] and for *dmr, sp91, sp89* and *ov7* [3], for the time being I believe it can be quite misleading because we cannot be confident yet that the data has been properly cleaned of secondary backgrounds, foregrounds and instrumental systematics to reveal the underlying primary anisotropies. Until then, band-power figures such as Fig.4. should be our guide to the evolving progress towards a primary $\mathcal{C}_\ell$ spectrum, and the theory of fluctuation generation underlying it.

## 4. DMR and FIRS

In this section, I describe some Bayesian results on the *firs* and first year *dmr* maps which use all aspects of the maps simultaneously and so are highly sensitive to all components in it. Since we wish to extract only the primary signals, filtering unwanted residual signals is essential, but there is some danger in doing so. I compare my unfiltered and filtered results with those obtained by the *dmr* and *firs* teams, whose techniques also employ filtering of one sort or another.

*4.1. Signal-to-noise Eigenmodes for Maps*

A full Bayesian analysis of maps requires frequent inversion and determinant evaluations of $N_{pix} \times N_{pix}$ correlation matrices, the sum of all $C_{Tpp'}$ in the theoretical modelling plus the pixel-pixel observational error matrix $C_{Dpp'}$. To facilitate this, I expand the pixel values $(\Delta T/T)_p$ into a basis of "signal-to-noise" eigenmodes for the maps in which the transformed noise and transformed (wanted) theoretical signal we are testing for do not have mode-mode correlations, *i.e.*, which are orthogonal. This can always be done, no matter what the experiment. Complications are associated with the removal of averages, dipoles *etc.* and the existence of secondary signals in the data, both of which do couple the modes. A model for the various contributions that make up the observed data is then

$$\xi_k = \sum_{p=1}^{N_{pix}} (RC_D^{-1/2})_{kp}(\Delta T/T)_p = s_k + (1+r)n_k + c_k + \mathrm{res}_k \ , \ k = 1, \ldots, N_{pix} \quad (38)$$

$$\langle n_k n_{k'} \rangle = \delta_{kk'} \ , \ \langle s_k s_{k'} \rangle = \mathcal{E}_{TRk}\delta_{kk'} = \left( RC_D^{-1/2} C_T C_D^{-1/2} R^\dagger \right)_{kk'} ,$$

$$\langle \mathrm{res}_k \mathrm{res}_{k'} \rangle = \mathcal{R}_{kk'} \ , \ \langle c_k c_{k'} \rangle = \mathcal{K}_{kk'} \ . \quad (39)$$

Because the transformation $C_D^{-1/2} C_T C_D^{-1/2}$ has dimension (*theory variance/pixel error*$^2$), I call the $\mathcal{E}_{TR,k}$ $S/N$-eigenvalues. I sort the modes in order of decreasing $\mathcal{E}_{TR,k}$, so low $k$-modes probe the theory in question best.

Because there are an equal number of eigenmodes as pixels, this new expansion is a complete (unfiltered) representation of the map. With uniform weighting and all-sky coverage, the eigenmodes are just the independent $Re(a_{\ell m})$ and $Im(a_{\ell m})$, with the lowest $\ell$ having the highest $\mathcal{E}_{TR,k}$, hence $k \sim (\ell+1)^2$. With Galactic cuts followed by dipole removals, and especially with inhomogeneous pixel coverage – a bigger issue for *firs*, but important for *dmr* too – they are complicated and theory-dependent. However, the high $S/N$-modes are indeed the ones that involve large scale pixel linear combinations, while the low $S/N$-modes typically involve positive and negative contributions from nearby pixels that are not sensitive to large scale structure in the maps, but are quite sensitive to physics inside the beam, whether from systematic effects or true white noise on the sky. This suggests this can be an ideal set for filtering. Filtering using $S/N$-modes has a long history in signal processing where it is called the Karhunen-Loeve method [67].

There is an arbitrary average and dipole that can be added to the maps, $c_k$, which I take to be described by a Gaussian with very wide width, *i.e.*, a uniform prior probability. In $S/N$-space, this contribution has off-diagonal correlations which affects small $k$. The residuals are modelled by an excess pixel noise with an amplification factor $r$, which soaks up a significant part of the excess I observe in the data, and an unknown component denoted by $\mathrm{res}_k$. Of course without identifying it, we do not know its correlation matrix $\mathcal{R}$, but the data itself can tell us something about its structure.

The sum of $|\xi_k|^2$ over bands in $S/N$-space defines a $S/N$ power spectrum which is easy to interpret because the modes are basically independent of each other, but have the disadvantage of depending upon the theory being tested for. In [7], I showed how *dmr* and

| Table 1: $\langle \mathcal{C}_\ell \rangle_{dmr}^{1/2}/10^{-5}$ as a function of $dmr$ map ($n_{\Delta T} + 3 = 1$) | | | | | | |
|---|---|---|---|---|---|---|
| 53a+b | 53a+b(7°) | 53a−b | 90a+b | 90a−b | 31a+b | firs |
| 1.03±.15 | 0.99±.14 | 0.30±.30 | 1.08±.21 | 0.00±.30 | 0.89±.35 | 1.09±.26 |

| Table 2: $\langle \mathcal{C}_\ell \rangle_{dmr,firs}^{1/2}/10^{-5}$ as a function of $n_s \equiv n_{\Delta T} + 3$ | | | | | | |
|---|---|---|---|---|---|---|
| 3.0 | 2.5 | 2.0 | 1.5 | 1.0 | 0.5 | 0.0 |
| *dmr*53a+b | | | | | | |
| 1.10±.10 | 1.07±.11 | 1.05±.12 | 1.02±.14 | 1.03±.15 | 1.05±.19 | 1.07±.22 |
| *firs* | | | | | | |
| 1.38±.26 | 1.33±.25 | 1.24±.25 | 1.15±.26 | 1.09±.26 | 1.05±.29 | 1.04±.32 |

| Table 3: $\langle \mathcal{C}_\ell \rangle_{dmr,firs}^{1/2}/10^{-5}$ as a function of $k_{cut}$ | | | | | | |
|---|---|---|---|---|---|---|
| $16^2$ | $14^2$ | $13^2$ | $10^2$ | $7^2$ | $5^2$ | $4^2$ |
| *dmr*53a+b, $n_{\Delta T} + 3 = 1$ | | | | | | |
| 1.03±.15 | 1.02±.15 | 1.02±.15 | 0.94±.14 | 0.85±.14 | 0.87±.18 | 0.98±.24 |
| *dmr*53a+b, $n_{\Delta T} + 3 = 2$ | | | | | | |
| 1.06±.12 | 1.05±.12 | 1.05±.12 | 0.97±.12 | 1.02±.15 | 1.05±.21 | 1.21±.29 |
| *firs*, $n_{\Delta T} + 3 = 1$ | | | | | | |
| 1.11±.27 | 1.07±.26 | 1.05±.26 | 1.00±.24 | 0.99±.25 | 0.89±.26 | 1.01±.32 |
| *firs*, $n_{\Delta T} + 3 = 2$ | | | | | | |
| 1.27±.24 | 1.21±.23 | 1.17±.23 | 1.08±.23 | 1.12±.27 | 1.14±.32 | 1.14±.38 |

| Table 4: $\langle \mathcal{C}_\ell \rangle_{dmr,firs}^{1/2}/10^{-5}$ from the *dmr* and *firs* teams | | | | |
|---|---|---|---|---|
| dmr1: S92 | dmr1: W94 | dmr2: B94 | dmr2: B94 ($n_{\Delta T} + 3 = 1.59$) | firs: G94 |
| 0.97±.28 | 0.97±.16 | 1.00±.10 | $1.02^{+.43}_{-.27}$ | 1.08±0.3 |

*firs* spectra for both $n_{\Delta T} + 3 = 1$ and 2 reveal excess power at low $S/N$ (high $k$) in the data, which neither beam-filtered theory can account for. Adding the constant $r \neq 0$ pixel noise enhancement is quite a good model for the excess $S/N$ power at the high $k$ end of the *firs* data, and not as good a model for the *dmr* 53a+b GHz data, but for *dmr* the map-dependent preferred $r$ is significantly smaller than for *firs*. For both $S/N$ power spectra, there is an excess at $k$ about $14^2$ that the power law theories cannot account for, and that plagues the statistical analysis of $n_{\Delta T}$. The nature of the modes and of the residual can also be probed by testing for its angular structure with correlation functions for $S/N$-filtered maps with high pass and low pass filters (as described in Fig.7.).

    The first step in the Bayesian method is the construction of a joint likelihood function in $\langle \mathcal{C}_\ell \rangle_{dmr,firs}^{1/2}$, $n_{\Delta T}$ and $r$. For given $n_{\Delta T}$, these reveal a strong maximum in $\langle \mathcal{C}_\ell \rangle_{dmr,firs}^{1/2}$. Integrating over $r$ (marginalizing it) allows one to construct $n_{\Delta T}$-$\langle \mathcal{C}_\ell \rangle_{dmr,firs}^{1/2}$ contour maps. Results for the 50% Bayesian probability value of $\langle \mathcal{C}_\ell \rangle_{dmr}$, with 'one-sigma' error bars determined by using the 84% and 16% Bayesian values are given in Tables 1-3: Table 1 gives band-powers for fixed $n_{\Delta T} + 3 = 1$ for various map combinations, which show very good agreement between the 53, 90 and 31 GHz *dmr* a+b maps and the *firs* map, and with no discernable signal in the *dmr* a−b maps; Table 2 shows that the derived band-power is very insensitive to the index $n_{\Delta T}$; Table 3 shows (for $n_{\Delta T} + 3 = 1$ and 2) that the derived band-power is remarkably insensitive to the signal-to-noise cut. $k_{cut}$ gives the number of modes kept, to be compared with 928 modes for the *dmr* maps (5.2° pixel size chosen, Galactic latitude cut of 25°) and 1070 for the *firs* map (2.6° pixel size chosen). In Table 4, I compare my results with the original first year *dmr* number [32], a later update [68], the new result using the combined first and second year data [63], derived for $n_{\Delta T} + 3 = 1$ and also for the most probable index, and the recent *firs* team result using the correlation function [69].

| Table 5: $n_{\Delta T} + 3$ for $dmr$53a+b and $firs$ | | | | |
|---|---|---|---|---|
| dmr | dmr(7°) | dmr, $k \leq (14)^2$ | dmr, $k \leq (13)^2$ | firs |
| $2.0^{+0.4;+0.7}_{-0.4;-1.0}$ | $1.7^{+0.4;+0.7}_{-0.4;-0.9}$ | $1.5^{+0.6;+1.4}_{-0.7;-1.0}$ | $1.0^{+1.1;+1.6}_{-0.7;-1.0}$ | $1.8^{+0.6;+0.9}_{-0.8;-1.3}$ |

| Table 6: $n_{\Delta T} + 3$ $dmr$ Team, First Year, and $firs$ Team | | |
|---|---|---|
| dmr1: S92(CF) | dmr1: W94(PS) | firs(CF) |
| $1.15^{+0.45}_{-0.65}$ | $1.69^{+0.45}_{-0.52}$ | $1.0^{+0.4}_{-0.5}$ |

| Table 7: $n_{\Delta T} + 3$ $dmr$ Team, Second Year Indices | | | |
|---|---|---|---|
| dmr2: B94(CF) | dmr2: W94(PS,3-19) | dmr2: W94(PS,3-30) | dmr2: G94 (3-30fil) |
| $1.59^{+0.49}_{-0.55}$ | $1.46^{+0.39}_{-0.44}$ | $1.25^{+0.4}_{-0.45}$ | $1.10^{+0.32}_{-0.32}$ |

I conclude that the band-power is very robust and well determined, at nearly the same level for all of the $dmr$ maps and for the $firs$ map, and largely independent of $n_s$. By contrast, the quadrupole power $\mathcal{C}_2$ (or equivalently $Q_{rms,PS} = 2.726K\sqrt{(5/12)\mathcal{C}_2}$) varies considerably, being related to the band-power by

$$\mathcal{C}_2 = 10^{-10}\left(\frac{Q_{rms,PS}}{17.6\mu K}\right)^2 \approx \langle\mathcal{C}_\ell\rangle_{dmr}\, e^{-\alpha(n_s-1)(1+0.3(n_s-1))}, \quad \alpha_{dmr} = 0.73,\ \alpha_{firs} = 1.1 \quad (40)$$

The anisotropy colour index $n_{\Delta T} = n_s - 3$ is another story. Integrating over $\langle\mathcal{C}_\ell\rangle^{1/2}_{dmr,firs}$ as well as $r$ gives the probability distribution for $n_{\Delta T}$. Table 5 gives colour indices for $dmr$53a+b, both unfiltered and with $k_{cut}$ in the region where the $S/N$ power spectrum reveals the excess that single $n_{\Delta T}$ laws can't explain. Maximum likelihood values with Bayesian 1-sigma and 2-sigma errors are shown: the index is determined with significantly less precision than the band powers are. The $k \leq (16)^2$ result is similar to the unfiltered result. As expected, sharp $S/N$-filtering does lower the index as the residual seen in the $S/N$ power spectra plots in [7] is chopped off. The signal-to-noise for modes with $k \geq 13^2$ is below 0.3, whereas the lowest 50 $dmr$ and 20 $firs$ modes have signal-to-noise in excess of unity. The non-diagonal part of $C_D$ does not change the Bayesian 50% value in Table 1, and increases the most likely band-power by 1.6%. The calculations used the $dmr$ beam and digitization and pixelization corrections advocated in [70]. To check that a faulty beam is not the problem, I tried a smaller 7° beam with no pixelization corrections, which lowers the value by 0.3. (The corresponding band power is listed in Table 1). A physical $n_{\Delta T} = 0$ white noise source as well as a long wavelength $n_{\Delta T} \approx -2$ would work quite well, but I haven't explored it yet, although allowing $r$ to float as a function of cut mimics the effect: that strategy robs the scale-invariant band-power by $\sim 20\%$ to feed an increased $r$ [7]. Tables 6 and 7 give various results obtained by the $dmr$ and $firs$ teams, using correlation function (CF) analysis (S92 [32], B94 [63] and firs [69]), quadratic power spectrum estimation (PS, W94 [71], with index determination over the $\ell$ range shown) and using a linear multipole filtering method (G94 [72]).

4.2. *Sharp S/N-filtering and Weiner-filtering of Maps*

I have found sharp $S/N$-filtering preferable to smooth $S/N$-filtering for statistical analysis, but some examples of the statistical use of smooth $S/N$-filtering were given in [4, 7]. One immediate byproduct of the $S/N$ eigenmode expansion is optimal or Weiner filtering, very useful for constructing maps cleaned of noise (Figs.5.,6.) to show robustness of structure from map to map. In Ref. [73], Bunn *et al.* have independently applied Weiner filtering, to the 'reduced galaxy' $dmr$ map, a linear combination of the 3 frequency maps. Because the Weiner filter changes with both $n_{\Delta T}$ and $\langle\mathcal{C}_\ell\rangle_{dmr,firs}$, I have not found it to be useful for the statistical analysis of either band-power or spectrum-colour.

Given observations $\bar{\xi}_k$, the mean value and variance matrix of the desired signal $s_k$ are (see *e.g.*, Appendix C of [19])

$$\langle s_k | \bar{\xi} \rangle = \sum_{k'} \left\{ \mathcal{E}_{TR} [\mathcal{E}_{TR} + (1+r)^2 + \mathcal{R} + \mathcal{K}]^{-1} \right\}_{kk'} \bar{\xi}_{k'} \qquad (41)$$

$$\langle \Delta s_k \Delta s_{k'} | \bar{\xi} \rangle = \left\{ \mathcal{E}_{TR} [\mathcal{E}_{TR} + (1+r)^2 + \mathcal{R} + \mathcal{K}]^{-1} [(1+r)^2 + \mathcal{R} + \mathcal{K}] \right\}_{kk'}. \qquad (42)$$

The mean field $\langle s_k | \bar{\xi} \rangle$ is the maximum entropy solution. The operator multiplying $\bar{\xi}_k$ is the Weiner filter. With no constraints and no extra residual, it is just $\mathcal{E}_{TR,k}/((1+r)^2 + \mathcal{E}_{TR,k})$. The fluctuation of the signal about the mean is $\Delta s_k = s_k - \langle s_k | \bar{\xi} \rangle$, a realization of which is found by multiplying a vector of $N_{pix}$ independent Gaussian random numbers by the square root of the variance matrix ($[\mathcal{E}_{TR,k}/(1+\mathcal{E}_{TR,k})]^{1/2} (1+r)$ with no constraints).

The Weiner filter depends upon the overall amplitude we think the theory signal has, parameterized by $\langle \mathcal{C}_\ell \rangle_{dmr,firs}$. I use the maximum likelihood values of $\langle \mathcal{C}_\ell \rangle_{dmr,firs}$ in the following. When the noise is large, as it is for these maps, it is the higher $\ell$ power that is preferentially removed by optimal filtering. Thus the theoretical fluctuation $\Delta s_k$ would have to be added to the Weiner-filtered maps of Figs.5.,6. to have a realistic picture of the sky given the data and the theory. That is the maps are too smooth, and more so for the noisier 90 GHz map than the 53 GHz map. Still it is very encouraging that the same large scale contour features persist in both maps. In the even noisier 31 GHz map, only a hint of the features persist visually because the $S/N$ filtering is so strong, but it is evident in the correlation function (Fig. 8.).

The distinguishing feature of the filtering procedures used here and in [72] is that they act linearly on the pixel amplitudes. Although using combinations, $(\Delta T/T)_p (\Delta T/T)_{p'}$, of pixel pairs is more complicated for statistical analysis, very useful forms of filtering become straightforward with these quadratic statistics. The correlation function is the classic example. The pixel-error enhancement model for the residual (parameter $r$) only contributes to the zero angle bin of the correlation function. The excess residual (beyond pixel-enhancement) resides between $k$ about $16^2$ to $13^2$. Fig. 7.(a) shows the *dmr* correlation function for modes above and below these values. Outside of the $\sim \theta_{fwhm}$ beamsize, $C(\theta)$ is nearly zero for the low $S/N$ modes and is nearly the same as $C(\theta)$ with no cut for the high $S/N$ modes. The same story basically holds true for *firs*, as can be seen in (d), except the deviation extends to $\sim 2\theta_{fwhm}$. (Note that what is uncorrelated in the full map *should* have some residual extended correlation in the filtered map). The other panels show the correlation function for a map realization of pure signal (b,e) and pure noise (c,f). For the pure noise simulation, the cut does matter, while for the pure theory simulation, $\mathcal{E}_{TRk}$ is so small above $16^2$ that it doesn't. I conclude from this that because $C(\theta)$ filters small-angle systematic (or physical) effects (if the first few angular bins are downweighted), $n_{\Delta T}$ determined this way provides a good indicator of the angular colour of the large-angle sky. Although bins within $\theta_{fwhm}$ are usually included, they are not overly weighted relative to larger angle bins.

In Fig.(8.), I contrast the highly noisy correlation functions (showing the error bars) for the unfiltered maps with $C(\theta)$ obtained for the Weiner-filtered maps. They reveal a number of remarkable points: (1) the optimally filtered correlation functions are effectively $n_{\Delta T}$-independent (shown are the $n_{\Delta T} + 3 = 1, 2$ cases for *dmr*53a+b and *firs* in (a) and (c)). Of course this does not address how statistically significant the optimally-filtered correlation function is for a given value of $n_{\Delta T}$. (2) The optimally filtered a-b *dmr* maps have zero $C(\theta)$. (3) The optimally-filtered a+b *dmr* $C(\theta)$'s shown in (f) on an expanded scale agree rather nicely. This is especially gratifying for the very noisy 31a+b map. The depth of the negative trough at $\sim 50°$ damps with increasing map noise, consistent with Weiner-filtering getting rid of the higher $\ell$ contributions. The next step is to apply the $S/N$ eigenmode method to the combined first and second year *dmr* data. This should be better behaved statistically, although [72] point to a continuing problem in the $dmr$53a+b map for $14 \lesssim \ell \lesssim 19$ modes, just where I have found anomalies that cannot be fit by a simple power law that are largely responsible for the steep indices.

## 5. Large Scale Structure Constraints

### 5.1. Parameterizing the Shape of the Density Power Spectrum

Shape constraints derived from large scale structure observations rely on the assumption of a uniform biasing and linearity on large scales, with different biasing parameters for the different objects observed. If so, the galaxy-galaxy and cluster-cluster power spectra and correlation functions directly reveal the underlying linear density power spectrum, $\mathcal{P}_\rho(k)$. It is useful to discuss these observations in terms of physically-motivated parameters that define the curvature of $\mathcal{P}_\rho(k)$, the initial scalar spectral index $n_s(k)$ and the scale of the horizon at redshift $\Omega_{nr}/\Omega_{er}$ when the density in nonrelativistic matter, $\Omega_{nr}\bar{a}^{-3}$, equals that in relativistic matter, $\Omega_{er}\bar{a}^{-4}$,

$$k_{Heq}^{-1} = 5\,\Gamma^{-1}\,\mathrm{h}^{-1}\mathrm{Mpc}\,,\quad \Gamma = \Omega_{nr}\mathrm{h}\,[\Omega_{er}/(1.69\Omega_\gamma)]^{-1/2}\,. \tag{43}$$

In [1], we adopted the now-widely-used functional form:

$$\mathcal{P}_\rho(k) \propto k^{3+n_s(k)}\left\{1+[ak+(bk)^{3/2}+(ck)^2]^\nu\right\}^{-2/\nu}, \tag{44}$$
$$(a,b,c) = (6.4, 3.0, 1.7)\Gamma^{-1}\mathrm{h}^{-1}\mathrm{Mpc}\,,\nu=1.13\,.$$

For $\Gamma = 0.5$, this accurately fits the linear power-spectrum of the standard CDM model with $\Omega_{nr} = 1$, h = 0.5 and $\Omega_B \approx 0.05$ [56]. The oft-used $\Omega_B \to 0$ form given in [19], Appendix G, is best fit by $\Gamma = 0.53$.

To fit the APM angular correlation function using a power spectrum for galaxies described by eq.(44) requires $0.15 \lesssim \Gamma \lesssim 0.3$ [1] for $n_s = 1$ and $0.2 \lesssim n_s \lesssim 0.6$ for $\Gamma = 0.53$ [2, 74]. A recent estimate of $\Gamma$ using power spectra from redshift surveys as well as from the APM data suggests $\Gamma \approx 0.25$ fits best [75]. Fig.1. compares the COBE-normalized $n_s = 1, \Gamma = 0.5$ linear density power spectrum with an $n_s = 1, \Gamma = 0.25$ and an $n_s = 0.6, \Gamma = 0.5$ spectrum.

To lower $\Gamma$ into the 0.15 to 0.3 range one can: lower h; lower $\Omega_{nr}$; or raise $\Omega_{er}$ (= $1.69\Omega_\gamma$ with the canonical three massless neutrino species present). Low density CDM models in a spatially flat universe (i.e. with $\Lambda > 0$) lower $\Omega_{nr}$ to $1 - \Omega_\Lambda$. CDM models with decaying neutrinos raise $\Omega_{er}$ [20, 76]: $\Gamma \approx 1.08\Omega_{nr}\mathrm{h}(1+0.96(m_\nu \tau_d/\mathrm{kev\,yr})^{2/3})^{-1/2}$ where $m_\nu$ is the neutrino mass and $\tau_d$ is its lifetime. Decaying neutrino models have the added feature of a bump in the power at subgalactic scales to ensure early galaxy formation, a consequence of the large effective $\Omega_{nr}$ of the neutrinos before they decayed.

Generally, more scales are needed to characterize the spectrum than just $k_{Heq}$, $e.g.$, the free-streaming scale for light neutrinos. In Hot/Cold hybrid models, there is a stable light neutrino of mass $m_\nu$ contributing a density $\Omega_\nu = 0.3(m_\nu/7.2\,\mathrm{ev})(2\mathrm{h})^{-2}$, combining with the CDM and baryon densities to make a total $\Omega_{nr} = 1$. A $\Gamma$-shape is not a very accurate representation of the entire spectrum, dropping from about 0.5 for small $k$ to $\Gamma \sim 0.22(\Omega_\nu/0.3)^{-1/2}$ over the band $0.04$-$2\,\mathrm{h}^{-1}\mathrm{Mpc}$ of relevance to $w_{gg}(\theta)$ calculations [20, 77].

To lower $n_s$, one can invoke one of the inflation models of § 2. utilizing a deceleration parameter $q \approx -(n_s+1)/2$ or, for natural inflation, the curvature in $\ln H$ away from the peak of the potential, $\frac{m_P^2}{4\pi}\partial^2 \ln H/\partial\phi^2 \approx (n_s-1)/2$.

Fig. 9. shows $n_{\rho,eff} \equiv d\ln\mathcal{P}_\rho(k)/d\ln k$ for the models we have been discussing and compares it with the indications from observation, the $0.15 \lesssim \Gamma \lesssim 0.3$ 'box' extending from $k^{-1} = 100\,\mathrm{h}^{-1}\mathrm{Mpc}$ downward into the nonlinear region (indicated roughly by the intersection of the $\gamma = 1.8$ line with 'box'.) Within the 'box', the preferred 0.25 value is also drawn. It can be seen that the effective index function for the standard CDM model is well outside of the box. So also are the hot/cold hybrid models shown (with a 2.4 and 7.2 ev $\nu$), although the $w_{gg}(\theta)$ doesn't look too bad. The model with $n_s$ tilted to 0.5 falls within the 'box', as does a model with the lesser tilt, 0.7, but with $H_0$ lowered as well, to 40. Of course, any $n_s = 1$ $\Gamma$-model with $\Gamma \sim 0.25$, such as one with a nonzero cosmological constant or a decaying $\nu$ also falls within the box. The large circle shows the rough value for the slope indicated by the X-ray temperature distribution function of clusters and the smaller one, by the X-ray luminosity distribution function, although much can go wrong in the interpretation of the latter [78].

As we have seen, for inflation models we expect that there will be some tilt, so probably we can relax the $\Gamma$ requirement. It is interesting to see what we would have to do to mimic a $\Gamma = 0.2$ shape with an initial spectral index variation if we were to assume a standard CDM model. The dotted $n_s(k)$ curve in Fig. 9. shows that radically-broken scale invariance is needed that changes the tilt from $\gtrsim 0.9$ beyond $\sim 200\,h^{-1}$Mpc to $\sim 0.5$ at $\sim 20\,h^{-1}$Mpc.

### 5.2. Relating the Cluster-amplitude $\sigma_8$ and the dmr Band-power

Apart from the shape parameters for $\mathcal{P}_\rho(k)$, there is also an overall amplitude parameter, which we now take to be $\langle \mathcal{C}_\ell^{(s)} \rangle_{dmr} = \langle \mathcal{C}_\ell \rangle_{dmr}/(1+\tilde{r}_{ts})$, eq.(31). Pre-COBE it was taken to be $\sigma_8$, the rms (linear) mass density fluctuations on the scale of $8\,h^{-1}$Mpc, which translates to the mass of a rich cluster, $1.2 \times 10^{15}\Omega_{nr}\,((2h)^{-1}\,M_\odot$: the number of rich clusters is extremely sensitive to the value of $\sigma_8$. Cluster X-ray data implies $0.6 \lesssim \sigma_8 \lesssim 0.8$ for CDM-like $\Omega_{nr} = 1$ theories ([78], and references therein). We shall consider 0.7 as the target number and values below 0.5 as unacceptable. A higher value is better for $\Omega_\Lambda \neq 0$ [1, 79].

The relation between $\sigma_8$ and the dmr band-power is[4]

$$\Gamma - \text{law}: \quad \sigma_8 \approx 1.2 \frac{\langle \mathcal{C}_\ell \rangle_{dmr}^{1/2}}{(1+\tilde{r}_{ts})^{1/2}10^{-5}} \Omega_{nr}^{-0.77}(2(\Gamma - 0.03))\, e^{-2.63(1-n_s)} , \qquad (45)$$

$$\text{hot/cold}: \quad \sigma_8 \approx 1.1 \frac{\langle \mathcal{C}_\ell \rangle_{dmr}^{1/2}}{(1+\tilde{r}_{ts})^{1/2}10^{-5}} (1+0.55(\Omega_\nu/0.3)^{1/2})^{-1}\, e^{-2.63(1-n_s)} . \qquad (46)$$

This shows that for the observed band-power, the $n_s = 1$ CDM model gives $\sigma_8$ too high, 1.1, but this is a sensitive function of $n_s$, dropping to 0.5 at $n_s = 0.82$ with the standard gravity wave contribution or at $n_s = 0.7$ if there is no gravity wave contribution. For the decaying neutrino model with $n_s = 1$, to have $\sigma_8 > 0.5$, we need $\Gamma > 0.25$ or $m_\nu \tau_d < 7.5$ kev yr. For the hot/cold hybrid model, we need $\Omega_\nu \approx 0.37$ for $\sigma_8 = 0.7$. See also ref.[80].

Another way to constrain the amplitude of the power spectrum is from the redshift of galaxy formation. We do not know this, but it cannot be too low or we would get too few $z \sim 4$ quasars and too little neutral gas compared with that inferred using the damped Lyman alpha systems seen in the spectra of quasars. This suggests $2 \lesssim \sigma_{0.5} \lesssim 5$ or so, where $\sigma_{0.5}$ is the analogue of $\sigma_8$ but at a Galactic mass scale rather than a cluster mass scale. This is the Achille's heel of hot/cold hybrid models [20, 80]. It also leads to serious constraints on $n_s$ for $\Gamma = 0.5$: in [2], we showed that a fairly conservative estimate of the redshift of galaxy formation was $z_{gf} \approx 1.3\sigma_{0.5} - 1$, and that $\sigma_{0.5} \approx 6.2\sigma_8 e^{-(1-n_s)}$ for $\Gamma = 0.5$, leading to $n_s > 0.76$ with gravity waves, $n_s > 0.63$ without. With $\Gamma < 0.5$, the restrictions on the primordial spectral index from galaxy and cluster formation are even more severe. That is, only a little tilt is allowed.

The free-streaming velocities of galaxies allow another estimate of the amplitude. For the $\Gamma$-laws, the large scale rms bulk flows over 40 and 60 $h^{-1}$Mpc in km s$^{-1}$ are

$$\sigma_v(40\,h^{-1}\text{Mpc})/\sigma_8 = 317(2\Gamma)^{-0.72}\Omega_{nr}^{0.56}[1^{+0.348}_{-0.572}]\, e^{1.06(1-n_s)}\, c.f.\, 388\,[1 \pm 0.17] , \qquad (47)$$

$$\sigma_v(60\,h^{-1}\text{Mpc})/\sigma_8 = 252(2\Gamma)^{-0.79}\Omega_{nr}^{0.56}[1^{+0.348}_{-0.572}]\, e^{1.19(1-n_s)}\, c.f.\, 327\,[1 \pm 0.25] . \qquad (48)$$

---

[4] These fits were made in $n_s$ with $\Gamma$ fixed at 0.5, and in $\Gamma$ with $n_s$ fixed at 1, hence will be rough when both vary signficantly from these standard values.

The comparisons are to the 'POTENT' estimates from the data [21]. The $\pm 1\sigma$ range is large because of the 'cosmic variance' expected in the measurement of a single bulk streaming velocity, but even so, the constraints in $\Gamma - \Omega_{nr} - n_s$ space are notable. For example, the $\Gamma = 0.5$ model needs $n_s > 0.89$ with the typical gravity wave contribution and $> 0.72$ with none [2]; *i.e.,* in spite of the fact that tilted spectra have more large scale power, hence for a given $\sigma_8$ would have larger scale flows, the normalization to *dmr* implies just the opposite, that the index must be very nearly scale invariant. In [1], we used a different version of the velocity constraint with (formally) lower error bars: $\sigma_8 \approx 1.0 \pm 0.24 \, \Omega_{nr}^{-0.56}(2\Gamma)^{0.2}$ for $n_s = 1$ (derived by relating redshift survey results to streaming velocities [22], but with the simplifying assumption of a linear amplification bias $\approx \sigma_8^{-1}$ for galaxies). This gives a $\sigma_8$ incompatibity with the COBE-estimated value for $n_s = 1$ $\Lambda > 0$ models with $\Gamma \sim 0.25$. The peculiar velocity data relies on having spatially-independent and accurate distance indicators (*e.g.,* the empirical Tully-Fisher relation between luminosity and rotation velocity in spiral galaxies). How seriously we take these constraints depends upon how reliable we think the indicators are – a subject of much debate.

*If the shape of the density power spectrum over the LSS band is now considered to be known, then this restricts the range of inflation and dark matter models considerably, especially when combined with the COBE anisotropy level. Whether the solution is a simple variant on the CDM+inflation theme involving slight tilt, stable ev-mass neutrinos, decaying >kev-neutrinos, vacuum energy, low $H_0$, or some combination, is still open, but can be decided as the observations tighten, and, in particular, as the noise in the $\mathcal{C}_\ell$ figure subsides, revealing the details of the Doppler peaks, a very happy future for those of us who wish to peer into the mechanism by which structure was generated in the Universe.* With the explosion of literature on such a wide-ranging topic, my referencing is far from complete – apologies. This work was supported by NSERC at Toronto and the Canadian Institute for Advanced Research.

Fig. 1. Cosmic waveband probes. The bands of cosmic fluctuation spectra probed by LSS observations are contrasted with the bands that current CMB experiments can probe. The (linear) density power spectrum for the standard $n_s = 1$ CDM model, labelled $\Gamma = 0.5$, is contrasted with power spectra that fit the galaxy clustering data, one tilted ($n_s = 0.6, \Gamma = 0.5$) and the other scale invariant with a modified shape parameter ($n_s = 1, \Gamma = 0.25$). Biasing must raise the spectra up (uniformly?) to fit into the hatched $w_{gg}$ range and nonlinearities must raise it at $k \gtrsim 0.2 \text{h Mpc}^{-1}$ to (roughly) match the heavy solid ($\gamma = 1.8$) line.

Fig. 2. (a) COBE-normalized temperature power spectra: for a standard $n_s = 1$ CDM model with standard recombination, early reionization, a (dashed) tilted primordial spectrum with $n_s = 0.95$, with the gravity wave contribution shown, and a (dotted) high $H_0$ model with $\Lambda \neq 0$. Bandpowers with 10% error bars are shown for the tilted and untilted CDM models. (b) Filter functions for current experiments.

Fig. 3. How a CDM model normalized to COBE varies with resolution. The contours begin at $109 \mu K$ in the half-degree smoothing cases, $54.5 \mu K$ in the no-smoothing case, $27.3 \mu K$ in the all-sky aitoff projection map. SR denotes standard recombination, NR denotes very early reionization, so there is no Doppler peak. The hills and valleys in the 5° SR map are naturally smooth: mapping them will give a direct probe of the physics of how the photon decoupling region at redshift $\sim 1000$ damped the primary signal.

Fig. 4. Band-power estimates for the vintage Spring 1994 data, decribed in detail in the text. The lower panel is a closeup of the Doppler peak region.

Fig. 5. (a) shows a 100° diameter map centred on the North Galactic Pole of the full $dmr$ 53a+b data, while (b) shows it after the Weiner-filtering, assuming a $n_{\Delta T} + 3 = 1$ $\mathcal{C}_\ell$ spectrum. (c) and (d) show the same for the 90a+b GHz data. The contours are $\pm 20 n \mu K$, $n \geq 1$ for (a,c), $\pm 10 n \mu K$ for (b,d). Positive contours are heavier than negative ones.

Fig. 6. (a,b,d) are the South Galactic Pole versions of Fig.5. (c) shows the Weiner-filtered 53a+b map with $n_{\Delta T} + 3 = 2$ is very similar in structure to (b) with $n_{\Delta T} + 3 = 1$.

Fig. 7. Correlation functions for $dmr$ (a-c) and $firs$ (d-f) maps. (a,d) shows the correlation function for the data with $k \leq (16)^2$ (closed triangles) and $k \geq (16)^2$ (open circles and nearly zero, except within the beam). The ×'s denote no cut, and for (a) the short horizontal lines denote $k \leq (13)^2$. (b,e) are two simulated noiseless skies for $n_{\Delta T} + 3 = 1$ (open circles) and 2 (closed squares), with the same random number seeds. The $k \leq (16)^2$ correlation is identical, while the $k \geq (16)^2$ one has no correlation. (c,f) shows the correlation function for pure noise (×'s: no cut; solid triangles: $k \leq (16)^2$; open circles: $k \geq (16)^2$).

Fig. 8. Correlation functions for unfiltered maps (open circles, dotted error bars) are compared with those for the Weiner-filtered maps (open squares for $n_{\Delta T} + 3 = 1$) in (a)-(e). (a) and (c) also show $n_{\Delta T} + 3 = 2$ correlations (solid triangles). Maximum likelihood amplitudes for the band power were chosen in each case. (f) compares the $dmr$ Weiner-filtered correlations directly: 53 (open square), 90 (closed triangle), 31 (closed circle). The high noise maps are of course more filtered than the low.

Fig. 9. $n_{\rho,eff}(k)$ as a function of scale. The $0.15 \lesssim \Gamma \lesssim 0.3$ box is the target shape which one can get by adjusting cosmological parameters to ensure $k_{Heq}^{-1} \sim 15-30\,\mathrm{h^{-1}Mpc}$ for $n_s = 1$ or by having $n_s \approx 0.5$ over the region of large scale clustering observations for $\Gamma = 0.5$.